# Thermodynamics and thermoeconomics of cell division in presence of exogenous materials in nucleus

Bin Kang[*]

State Key Laboratory of Analytical Chemistry for Life Science, School of Chemistry and Chemical Engineering, Nanjing University, Nanjing, 210023, China

[*]E-mail: binkang@nju.edu.cn

Abstract

Cell division is an essential biological process, and regulation of cell division is of relevance for many important fields of biology and medicine. Introducing exogenous substances, such as nanoparticles, into the nucleus, has been experimentally studied to regulate the division of cells. Herein we considered this phenomenon from a general view of energetics. Through analyzing the thermodynamics during the cell division process, we investigated the optimal symmetry for cell division and the effect of nanoparticles on the energy barriers. The presence of nanoparticles inside cell nucleus might arrest cells before cytokinesis or other stages, thereby regulate the cell division.

Cell division is an essential biological process that is normally well-regulated by a complex of signaling networks,[1] yet sometime, the lost control over division leads to unlimited cell growth i.e. cancer.[2] Regulation of cell division is of relevance for many important processes of biology and medicine research. Usually, cell division process could be regulated by specific molecules that block the function of key molecules that are required for complete division, which is the common mechanism of most chemotherapeutic drugs.[3,4] Previous works have reported the possibility of regulating the cell division by exogenous nanoparticles targeted to cell nucleus.[5,6] In this letter, we attempted to explore the nature of this phenomenon from a fundamental view point of energetics. Through analyzing the thermodynamic evolution during cell division process, we investigated the optimal symmetry for cell division and the effect of nanoparticles on the energy barriers. Calculation showed that nanoparticles inside nucleus above threshold concentration would arrest cells before cytokinesis or other stages, finally regulate the cell division process.

We considered the cell division process as a series of thermodynamic events with different energy states (Figure 1). According to the experimental observed chromosome patterns and cell shapes, the whole cell division process could be divided to several typical steps.[7-11] Considering a cell as a open system, thus at equilibrium state, the differential form of free energy contributed by the surface potential of cell membrane and the chemical potential of chromosomes can be given as:

$$dA = -PdV - SdT + \sum_{i=1}^{i} \mu_i dN_i + \gamma dA_s$$
$$dG = VdP - SdT + \sum_{j=1}^{j} \mu_j dN_j$$
(1)

where $A$ is the Helmholtz free energy of cell membrane, $G$ is the Gibbs free energy of chromosomes, $S$ is the entropy, $P$ is the pressure, $V$ is the volume, $T$ is the temperature, $\mu_i$ and $\mu_j$ are the chemical potential of matter $i$ and $j$, $N_i$ and $N_j$ is the mole number of matter $i$ and $j$, $\gamma$ is the surface tension of cell membrane, $A_s$ is the surface area of cell membrane.

Cell division occurs at constant temperature of 37 °C,[1] and the pressure inside cells is a constant,[2] thus the cell is a $P$, $T$ constant system. During mitosis, the chromosome concentrated and the process of transcription and translation intermitted,[1,2,4] so the total mount of components inside cells can be considered unchanged. Under above assumption, the equation (1) can be approximated as:

$$dA = \gamma dA_s$$
$$dG = \sum_{j=1}^{j} \mu_j dN_j \qquad (2)$$

the chemical potential in equation (2) can be given as:

$$\mu_j = \mu_j^0 + RT \ln \frac{m_j}{(\rho V / M)} \qquad (3)$$

where $\mu_j^0$ is the chemical potential of pure component $j$, $R$ is the Boltzmann constant, $m_j$ is the mole number of matter $j$, $\rho$ is the intensity of chromosome mixture, $M$ is the average mole mass of mixture, so $m_j / (\rho V / M)$ is the mole fraction of matter $j$.

Normally, cell divides symmetrically,[1,2] the chromosomes are equally distributed in two daughter cells and the daughter cells are same in size. This complex process is regulated by spindle,[12] actin filament[13] and cell signaling.[14] From a thermodynamic point of view, we assume the symmetry of cell division relates to the changes of free energy. Generally considering one cell that undergoes an asymmetric division, the two daughter cells are different in cell size and chromosome numbers. If the radius of two daughter cells are $r_d(1)$ and $r_d(2)$ respectively, thus the cell volumes $V_d(1)$ and $V_d(2)$ are: $V_d(1) = \frac{4}{3}\pi r_d(1)^3$, $V_d(2) = \frac{4}{3}\pi r_d(2)^3$, and the cell membrane area $A_d(1)$ and $A_d(2)$ are: $A_d(1) = 4\pi r_d(1)^2$, $A_d(2) = 4\pi r_d(2)^2$. The total cell volume is constant during cell division $V_d(1) + V_d(2) = \frac{4}{3}\pi R^3$, so $r_d(2) = \sqrt[3]{R^3 - r_d(1)^3}$, thus the surface energy change before and after cell division is:

$$\Delta A = \gamma \Delta A_s = \gamma 4\pi R^2 \left\{ \left(r_d(1)/R\right)^2 + \left[1 - \left(r_d(1)/R\right)^3\right]^{\frac{2}{3}} - 1 \right\} \quad (4)$$

Equation (4) only considering the changes of surface energy, the parent cell trends to divide asymmetrically (Figure 2a). However, for real cell division, the chemical potential of chromosomes must also be considered. Firstly, we consider the situation that two daughter cells have different chromosome number but equal cell size. The total mole number of chromosomes in parent cell is defined as $N$ and their volume is $V_1$, after division, one daughter cell gets chromosome of $xN$, thus another cell gets chromosome of $(1-x)N$, the chromosome volume of two daughter cells are $V(1)$ and $V(2)$. From equation (3), the chemical potentials of chromosomes in two daughter cells are:

$$\mu_d(1) = \mu^0 + RT \ln \frac{xm}{(\rho V(1)/M)}$$
$$\mu_d(2) = \mu^0 + RT \ln \frac{(1-x)m}{(\rho V(2)/M)} \quad (5)$$

after cell division, the Gibbs free energy changes can be given as:

$$\Delta G = \sum_{j=1}^{j} \mu_{d,j}(1) dN_j + \sum_{j=1}^{j} \mu_{d,j}(2) dN_j - \sum_{j=1}^{j} \mu_j(1) dN_j \quad (6)$$

combine equation (3) and (5) into equation (6):

$$\Delta G = NRT \left[ \left(x \ln x + (1-x)\ln(1-x)\right) + x \ln \frac{V(2)}{V(1)} + \ln \frac{V_1}{V(2)} \right] \quad (7)$$

In equation (7), the chromosomes with number $xN$ have a volume $V(1)$, the chromosomes with number $(1-x)N$ have a volume $V(2)$, thus we can consider two situations: (1) If the concentration of chromosomes is a constant inside parent and daughter cells, i.e. $V(1)/V(2) = x/(1-x)$, $V(2) = xV_1$, there by the Gibbs free energy change $\Delta G = 0$.

(2) If the chromosome volume (nucleus volume) is a constant, i.e. $V(1) = V(2) = \frac{1}{2}V_1$, then the Gibbs free energy changes can be written as:

$$\Delta G = NRT\left[x\ln x + (1-x)\ln(1-x) + \ln 2\right] \qquad (8)$$

Equation (8) shows that the chromosomes in parent cell trends to separate equally into two daughter cells. In this situation, the change of surface energy $\Delta A$ is a constant (Figure 2b). Thus the optimization of free energy change results in symmetric cell division. We further considered another situation that two daughter cells have equal amount of chromosomes but different cell volume. If the parent cell has chromosome number of $N$, each daughter cell got $N/2$. Assuming the volume of chromosomes relates on the volume of cell, if the volume of one daughter cell is $yV_0$, where $V_0$ is the volume of parents cell, then the volume of chromosomes can be written as $\chi yV_0$, where $\chi$ is the volume ratio of chromosome and cell. For easy discuss, we can set $\chi = V_1/V_0$, where $V_1$ is the chromosome volume in parent cell. Thus the cell volume and chromosome volume of another daughter cell are $(1-y)V_0$ and $\chi(1-y)V_0$. The chemical potential of chromosomes in each daughter cell can be given as:

$$\mu_d(1) = \mu^0 + RT\ln\frac{m}{2y\chi V_0 \rho/M}$$
$$\mu_d(2) = \mu^0 + RT\ln\frac{m}{2(1-y)\chi V_0 \rho/M} \qquad (9)$$

combine equation (3) and (9) into equation (6), the Gibbs free energy change can be given as:

$$\Delta G = NRT\left(\frac{1}{2}\ln\frac{1}{y} + \frac{1}{2}\ln\frac{1}{1-y} - \ln 2\right) \qquad (10)$$

From equation (10), if the two daughter cells have equal chromosome numbers, they trend to have same cell volume to minimize the change of chemical potential. However, the equal division maximizes the change of surface energy (Figure 2c). Thus the change of total free energy during cell division could be expressed as $\Delta E_{Total} = \Delta G + \Delta A$. Combining equation (4) and (10) together, the $\Delta E_{Total}$ could be given as:

$$\Delta E_{Total} = NRT\left\{\left(\frac{1}{2}\ln\frac{1}{y} + \frac{1}{2}\ln\frac{1}{1-y} - \ln 2\right) + k\left[y^{\frac{2}{3}} + (1-y)^{\frac{2}{3}} - 1\right]\right\} \quad (11)$$

In equation (11), we defined a factor $k = \gamma 4\pi R^2 / NRT$ to relate the $\Delta G$ and $\Delta A$, and $y = (r_d(1)/R)^3$ is the volume ratio of one daughter cell with parent cell, namely symmetrical factor. Figure 3 shows the function of total free energy change $\Delta E_{Total}$ versus the symmetrical factor $y$ under the conditions of different $k$ value. The $k$ value is vital to the symmetry of cell division, cells trend to divide symmetrically at a small $k$ value (<3) but asymmetrically at a big $k$ value (>4).

We further analyzed the changes of free energy at each steps during cell division. Since cells divide symmetrically under normal conditions, thus for easy discussion, we consider following process as typical equal division with $k = 1$. In current model as showing in Figure 1, the surface areas of cell membrane at each steps were defined as $A_s(i)$, where $i = 1, 2...6$. Thus the change on surface area of cell membrane between two steps can be given as $\Delta A_s(i \rightarrow i+1)$, where $i = 1, 2...5$; then the Helmholtz energy changes are $\Delta A(i \rightarrow i+1) = \gamma \Delta A_s(i \rightarrow i+1)$, where $i = 1, 2...5$. From equation (1)-(3), the Gibbs energy changes could be expressed as following:

$$\Delta G(i \rightarrow i+1) = \sum_{j=1}^{j} \mu_j(i+1)dN_j - \sum_{j=1}^{j} \mu_j(i)dN_j = NRT\ln\frac{V_i}{V_{i+1}} \quad (12)$$

where $V_i$ and $V_{i+1}$ are the volume of chromosome at step $i$ and $i+1$, $i = 1, 2...5$. Then the changes of toal free energy from step $i$ to $i+1$ can be written as:

$$\Delta E_{Total}(i \rightarrow i+1) = \Delta A(i \rightarrow i+1) + \Delta G(i \rightarrow i+1) = \gamma \Delta A_s(i \rightarrow i+1) + NRT\ln\frac{V_i}{V_{i+1}} \quad (13)$$

where $i = 1, 2...5$.

Figure 4 shows the energy change profile at each stage during cell division (parameters: $R = 5 \mu m$, $d = r_3$, $h = \sqrt{2} r_4$, $l = \frac{1}{2} r_5$, $w = \frac{1}{4} r_5$). Cells need overcome two energy barriers to pass through division, the first barrier at metaphase is mainly contributed by the chemical potential, and the second barrier at cytolinesis is mainly contributed by the surface potential. Cells have various molecular motors involving spindle and actin filament to undergo chromosomes distribution and cytoplasm seperation.[12-14] This process consumes energy, and the consumed energy is provided by cell itself. From the view of thermodynamics, the energy that each cell can provide for division is finite. If the maximum amount of energy that cell can provide for division process can be defined as $E_{max}^0$, and the maximum amount of energy barrier that cells need to overcome in division process can be defined as $\Delta E_{Total}^{Max}$. Thus the success of cell division requires $E_{max}^0 \geq \Delta E_{Total}^{Max}$, which means that cells have enough energy to overcome the energy barriers to complete division. Assuming that the nanoparticles inside cell nucleus mix with chromosomes and they exhibit the same movement during division, then equation (13) can be evolved to:

$$\Delta E_{Total}(i \to i+1) = \gamma \Delta A_s(i \to i+1) + (N + N^*) RT \ln \frac{V_i}{V_{i+1}} \quad (14)$$

where $N^*$ is the equivalent mole number contributed by nanoparticles. From equation (14), there is a critical value $N_0^*$, when $N^* > N_0^*$, that $E_{max}^0 < \Delta E_{Total}^{Max}$. That is to say, the cells cannot provide enough energy to overcome the energy barrier when the amount of nanoparticles inside cell nucleus is higher than the threshold, thereby cells cannot complete the division process. The presence of nanoparticles inside cell nucleus caused the increase of energy barriers during division process. Once the energy barrier is more than the maximum cell energy, the complete cell division fails and cells are stuck before energy barriers. If the mole number of nanoparticles $N^*$ in nucleus is slightly higher than the critical mole number $N_0^*$, cells are arrested in cytokinesis, daughter cells fail to separate and finally fuse back to form a binucleate cell. If the mole number of nanoparticles in cell nucleus is much higher, cells might be arrested in earlier steps, such as metaphase.

In conclusion, our calculation demonstrated that the presence of exogenous materials like nanoparticles inside cell nucleus would change its energy barrier during cell division. The exogenous materials with concentration above threshold would arrested cells before cytokinesis or other stages, finally regulated the cell division. This result provided a general point of view about how thermodynamics affect the cell division, and offered a possible strategy for cell-division regulation by exogenous materials.

FIGURES

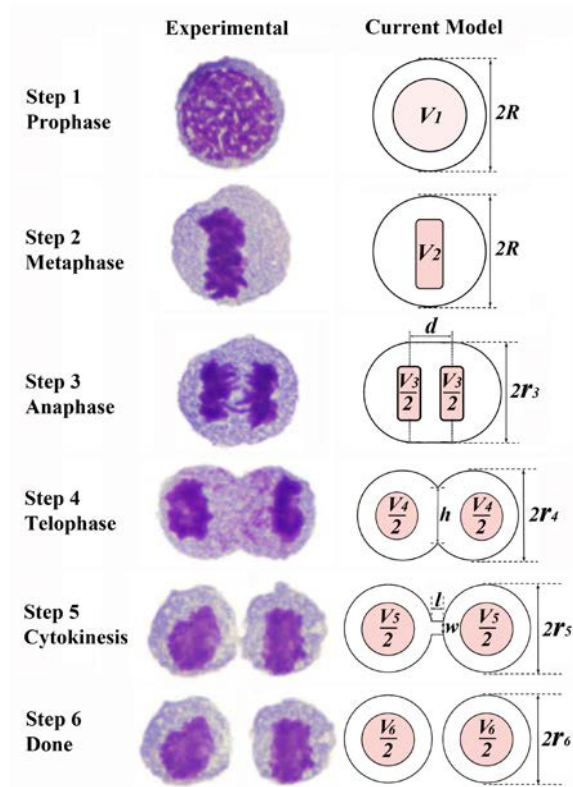

**Figure 1.** The experimental observation of mitosis and current proposed model. According to the movment of chromosomes and changes on cell shape, the whole division process could be devided to several typical steps.

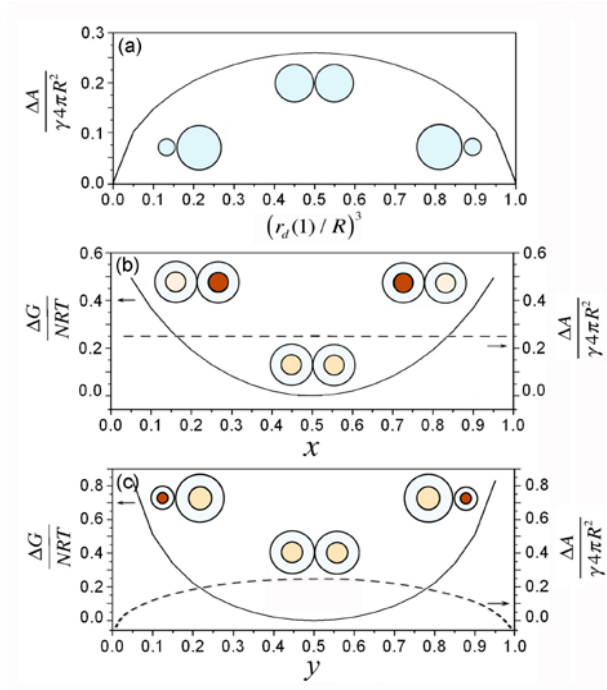

**Figure 2.** The changes of free energy versus the symmetry of cell division. (a) surface energy change versus daughter cell size, $(r_d(1)/R)^3$ is the volume ratio of one daughter cells with parent cell; (b) changes of chemical potential (solid line) and surface energy (dish line) versus chromosome distribution when two daughter cells with same size, $x$ is the ratio of chromosome number in one daughter cells with parent cell; (c) changes of chemical potential (solid line) and surface energy (dish line) versus daughter cell size when two daughter cells with same chromosomes, $y = (r_d(1)/R)^3$ is the volume ratio of one daughter cells with parent cell.

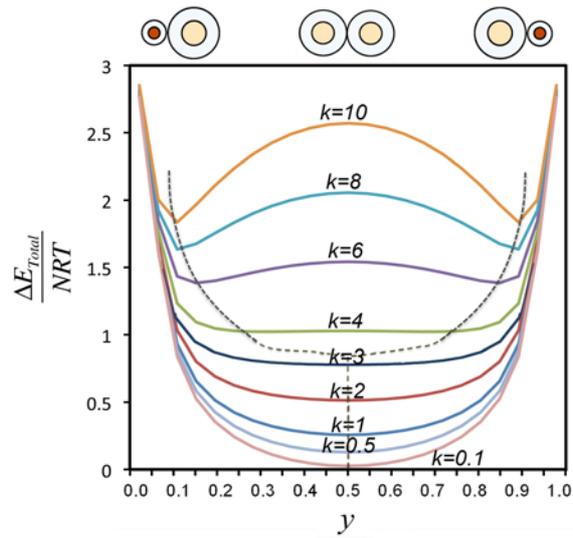

**Figure 3.** The changes of total free energy versus the symmetry of cell division at different $k$ value. The symmetrical factor $y$ indicates a symmetric division at $y = 0.5$, otherwise it is an asymmetric division. The dish lines indicate the minimium value points of free energy change, which means the optimal division symmetry.

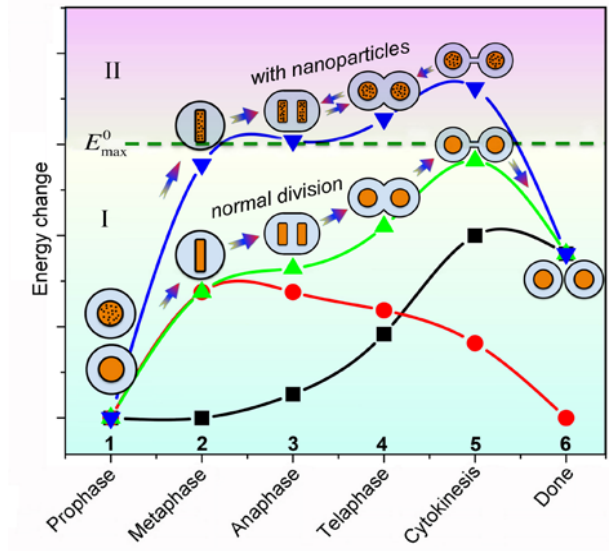

**Figure 4.** Profile of energy changes at each step during mitosis: ● chemical potential of chromosomes $\Delta G$; ■ surface potential of cell membrane $\Delta A$; ▲ total free energy $\Delta E_{Total}$ of normal cell division, and ▼ cell division with nanoparticles in nucleus. In region I that $\Delta E_{Total}^{max} < E_{max}^0$, cell divide normally; in region II that $\Delta E_{Total}^{max} > E_{max}^0$, complete cell division fails since the energy provided by cells is not enough to overcome the energy barrier.